\documentclass[twocolumn,prd,floatfix,nofootinbib]{revtex4}
\usepackage{amsmath}
\usepackage{graphicx}
\usepackage{dcolumn}
\usepackage{bm}
\usepackage{amssymb}
\usepackage{latexsym}
\usepackage{color}

\def\be{\begin{equation}}
\def\ee{\end{equation}}
\def\ba{\begin{eqnarray}}
\def\ea{\end{eqnarray}}

\newcommand{\bd}{\ensuremath{\overline{\textnormal{D3}}}}

\begin{document}

\title{CMB anomalies from an inflationary model in string theory}

\author{Zhi-Guo Liu$^{1,}$\footnote{Email: liuzhiguo08@mails.ucas.ac.cn}}
\author{Zong-Kuan Guo$^{2,}$\footnote{Email: guozk@itp.ac.cn}}
\author{Yun-Song Piao$^{1,}$\footnote{Email: yspiao@ucas.ac.cn}}

\affiliation{$^1$ School of Physics, University of Chinese Academy
of Sciences, Beijing 100049, China}

\affiliation{$^2$ State Key Laboratory of Theoretical Physics, Institute of Theoretical Physics, \\
Chinese Academy of Sciences, P.O. Box 2735, Beijing 100190, China}

\begin{abstract}

Recent Planck measurements show some CMB anomalies on large
angular scales, which confirms the early observations by WMAP. We
show that an inflationary model, in which before the slow-roll
inflation the Universe is in a superinflationary phase, can
generate a large-scale cutoff in the primordial power spectrum,
which may account for not only the power suppression on large
angular scales, but also a large dipole power asymmetry in the
CMB. We discuss an implementation of our model in string theory.

\end{abstract}


\maketitle


\section{Introduction }

Recently, the Planck collaboration has reported a hemispherical
power asymmetry in the CMB \cite{Ade:2013nlj}, which conformed the
result of WMAP,
but has better precision. Such asymmetry has also been found by
estimating the power spectrum in the two hemispheres by using the
quadratic maximum likelihood~\cite{Paci:2010wp}. In addition, the
Planck collaboration has also reported a power deficit in the
low-$l$ CMB power spectrum at $l\lesssim 40$ \cite{Ade:2013nlj}
with the statistical significance $2.5\sim 3\sigma$, which is not
concordant with the Planck bestfit model, although the data points
are still consistent well with the cosmic variance.

The Planck data have larger statistical significance than the WMAP
data, which makes the anomalies difficult to attribute the
foregrounds, e.g.\cite{Dai:2013kfa},\cite{Liddle:2013czu}. Thus it
seems that these anomalies should have an underlying and common
physical origin, which deserves to be considered seriously.


The CMB power asymmetry might be modeled as a dipole modulation of
the power \cite{Prunet:2004zy},\cite{Gordon:2005ai}, see also
\cite{Rath:2013yra}, which results from a superhorizon
perturbation crossing the observable Universe
\cite{Erickcek:2008sm},\cite{Lyth:2013vha}. This modulation can be
explained in light of the spatial change of the spectrum of
primordial curvature perturbation ${\cal R}$, \be {\cal
P}^{1/2}_{\cal R}(k,\mathbf{x})=\left(1+
A(k){\hat{\mathbf{p}}\cdot \mathbf{x}\over x_{\rm ls}}\right)
{\cal P}^{1/2}_{{\cal R}}(k), \ee where $\hat{\mathbf{p}}$ is the
unit vector of the dipole modulation direction, $x_{\rm ls}$ is
the distance to the last scattering surface, ${\cal P}_{{\cal
R}}(k)$ is the power spectrum with index $n_{\cal R}(k)$, and
$A(k)$ is the amplitude of modulation, which is
\cite{Lyth:2013vha},\cite{Liu:2013kea} \ba A(k) & = &
{|\nabla{\cal P}^{1/2}_{{\cal R}}(k,\mathbf{x})|\over {\cal
P}^{1/2}_{{\cal R}}}\,x_{\rm ls}\nonumber\\
& = & \left(1-\epsilon\right)\left[{n_{\cal R}(k)-1\over
2}\right]k_{\rm L} x_{\rm ls}{\cal P}^{1/2}_{{\cal R},{\rm L}}\,,
\label{tp}\ea where ${\cal P}_{{\cal R},{\rm L}}$ is the amplitude
of the power spectrum of the modulating mode $k_{\rm L}$, and
$\epsilon=-{\dot H}/H^2$. We have $(k_L x_{\rm ls}){\cal
P}^{1/2}_{{\cal R},{\rm L}}\lesssim 0.1$
\cite{Erickcek:2008sm},\cite{Lyth:2013vha},\cite{Namjoo:2013fka}.

In single field inflationary scenario, the spectrum $n_{
inf}-1\sim 0.04$ is almost scale invariant. Thus on large angular
scales the amplitude of the modulation is too small to fit the
observation \cite{Erickcek:2008sm},\cite{Lyth:2013vha}. In
addition, the almost scale invariance of the inflationary spectrum
also fails to explain the power deficit on large angular scales.





However, it could be observed that a large amplitude of the
modulation consistent with the observations actually requires the
breaking of the scale invariance of power spectrum on large
angular scales, while simultaneously such a breaking also helps to
explain the power suppression on corresponding scales,
e.g.\cite{Liu:2013kea}. In this angle of view, the anomalies on
large angular scales may be a hint of the pre-inflationary
physics, which might be relevant with the initial singularity,
e.g.\cite{Piao:2003zm},\cite{Dudas:2012vv}.

Here, we will show that an inflationary model, in which before the
slow-roll inflation the Universe is in a superinflationary phase,
can generate a large-scale cutoff in the primordial power
spectrum, which may account for not only the power suppression on
large angular scales, but also a large dipole power asymmetry in
the CMB.

It is generally thought that the pre-inflationary physics ought to
be controlled by a fundamental theory, e.g. string theory. How
embedding the inflationary scenario into string theory, has still
been a significant issue, which has been studied intensively, see
Ref.\cite{Burgess:2013sla}. Thus it is intriguing and might be
naturally expected that a stringy mechanism of inflation could
give the CMB anomalies on large angular scales,
e.g.\cite{Liddle:2013czu},\cite{Kanno:2013ohv} with string
landscape, and also \cite{Cicoli:2013oba},\cite{Pedro:2013pba}
with a fast-roll phase in fibre inflation \cite{Cicoli:2008gp}. We
will discuss an implementation of our model in string theory,
based on Refs.\cite{PRZ},\cite{DKV}.


\section{The modulating mode from a superinflationary phase}

We first will calculate the primordial perturbation generated in
such an inflationary model,
and identify the
corresponding modulating mode from a superinflationary phase.
The equation of the curvature perturbation $\cal R$ in momentum
space is 
\be u_k^{\prime\prime} +\left(c^2_s k^2-{z^{\prime\prime}\over
z}\right) u_k = 0, \label{uk}\ee  after $u_k \equiv z{\cal R}_k$
is defined, where $'$ is the derivative with respect to the
conformal time $\eta=\int dt/a$, $z\equiv {a\sqrt{2M_P^2\epsilon}/
c_s}$. We have $c_s^2=1$ for a canonical scalar field.

The Universe initially is in a superinflationary phase with
$\epsilon_{Pre-inf}\sim -{\cal O}(1)$, hereafter, it will get into
an inflationary phase with $\epsilon_{\rm inf}\ll 1$. We will
neglect the matching detail for simplicity. Thus in conformal
time, after adopting an instantaneous matching, we have \ba a &
\simeq & {a_0\over \sqrt{1-2{\cal H}_0\eta}}~,
\,\,\,for\,\,{\rm the }\,\,{\rm superinflation} \label{leq} \nonumber\\
& & {a_0\over 1-{\cal H}_0\eta}~, \,\,\, for \,\,{\rm the
}\,\,{\rm inflation}. \label{geq} \ea where $\eta<0$ in the
superinflationary phase and $\eta>0$ in the inflationary phase,
respectively, and $a=a_0$ for $\eta=0$ is set, ${\cal H}_0$ is the
comoving Hubble length at matching time $\eta=0$, which sets the
inflationary energy scale by $H_{\rm inf}={\cal H}_0/a_0$. Here,
$\epsilon_{Pre-inf}=-1$ is applied. In principle, other value with
$|\epsilon_{Pre-inf}| \gtrsim 1$ may be also used, which, however,
hardly alter the result qualitatively. The evolution of the
superinflationary phase with arbitrary $\epsilon<0$ and the
primordial perturbation generated have been studied earlier in
Ref.\cite{Piao:2004tq}. The case with $\epsilon\ll -1$ corresponds
to the slow expansion scenario of primordial universe, which has
been proposed earlier in Ref.\cite{Piao:2003ty} and investigated
in details in Ref.\cite{Piao:2010bi}.


When $k^2\gg {z^{\prime\prime}\over z}$, the perturbation is deep
inside its horizon, we have $u_k\sim {1\over \sqrt{2k}}
e^{-ik\eta}$. In the superinflationary phase, \be
{z^{\prime\prime}\over z}\simeq {3{\cal H}_0^2\over (1-2{\cal H}_0
\eta)^2}. \ee When $k^2\ll {z^{\prime\prime}\over z}$, the
solution of Eq.(\ref{uk}) is \be {u_k}=\sqrt{\pi(1-2{\cal
H}_0\eta)\over 8{\cal H}_0}H_1^{(1)}\left(-k\eta+{k\over 2{\cal
H}_0}\right)
 ,\ee where $H_1^{(1)}$ is the first-order Hankel function of the first kind.

In the inflationary phase, \be {z^{\prime\prime}\over z} \simeq
{2{\cal H}_0^2 \over (1-{\cal H}_0\eta)^2} .\ee When $k^2\ll
{z^{\prime\prime}\over z}$, i.e. $-k\eta +{k/{\cal H}_0}\ll 1$,
the solution of Eq.(\ref{uk}) is \ba
{u_k} & = & \sqrt{-k\eta +{k\over {\cal H}_0}}\nonumber\\
& & \left(C_1 H_{3/2}^{(1)}(-k\eta +{k\over {\cal H}_0})+C_2
H_{3/2}^{(2)}(-k\eta +{k\over {\cal H}_0})\right) \label{vki} ,\ea
where $H_{3/2}^{(1)}$ is the 3/2th-order Hankel function of the
first kind, $H_{3/2}^{(2)}$ is the 3/2th-order Hankel function of
the second kind, $C_1$ and $C_2$ are only dependent on $k$.

We require that all physical quantities continuously pass through
the matching surface. The continuity of curvature perturbation
gives \ba \label{c1} C_1 &=& { \frac{i\pi e^{-ik/{\cal
H}_0}}{16}\left(1-\frac{{\cal H}_0}{k}i\right)}
 \left[H^{(1)}_0(\frac{k}{2{\cal H}_0
})-H^{(1)}_2(\frac{k}{2{\cal H}_0})\right] \nonumber\\ & - &
\frac{\pi e^{-ik/{\cal H}_0}}{8}(1-\frac{2{\cal H}_0^2}{k^2}-\frac{2{\cal H}_0}{k}i)H^{(1)}_1(\frac{k}{2{\cal H}_0}),\\
C_2 &=& { -\frac{i\pi e^{ik/{\cal H}_0}}{16}\left(1+\frac{{\cal
H}_0}{k}i\right)} \left[H^{(1)}_0(\frac{k}{2{\cal H}_0
})-H^{(1)}_2(\frac{k}{2{\cal H}_0})\right] \nonumber\\ & - &
\frac{\pi e^{ik/{\cal H}_0}}{8}(1-\frac{2{\cal
H}_0^2}{k^2}+\frac{2{\cal H}_0}{k}i)H^{(1)}_1(\frac{k}{2{\cal
H}_0}), \label{c2} \ea where $H_0^{(1)}$ is the zeroth-order Hankel
function of the first kind and $H_2^{(1)}$ is the second-order
Hankel function of the first kind.

Thus the power spectrum of $\cal R$ is \be {\cal P}_{\cal R} =
{k^3\over 2\pi^2}\left|{u_k\over z}\right|^2={\cal P}_{\cal
R}^{\rm inf} {2\over \pi}k\left|C_1 -C_2\right|^2, \label{ps}\ee
where ${\cal P}_{\cal R}^{\rm inf} = {H_{\rm inf}^2\over 4 \pi^2
M_P^2 \epsilon_{\rm inf}}$ is that of the standard slow roll
inflation, which may has a slight red spectrum consistent with the
observation, and $C_1$ and $C_2$ are determined by Eqs.(\ref{c1})
and (\ref{c2}), respectively. The spectrum index of $\cal R$ is
$n_{\cal R}= n_{\rm inf}+{d\ln{(k|C_1 -C_2|^2)}\over d\ln k}$.



In Ref.\cite{Biswas:2013dry}, the perturbation from a
superinflationary phase was also calculated. However, it is
assumed that before the superinflationary phase there is a
nonsingular bounce appears, which is not required here.

Here, ${\cal H}_0$ is the comoving Hubble length at matching
surface $\eta=0$. The modulating mode corresponds to that on large
scales $k\ll {\cal H}_0$, which is generated during the
superinflationary evolution.
We may expand the Hankel functions in term of $k\ll {\cal H}_0$
and have \ba {\cal P}_{\cal R}^{k < {\cal H}_0} \simeq
{\cal P}_{\cal R}^{\rm inf}\frac{2k}{\pi {\cal
H}_0}\left(1+\frac{k^2}{12{\cal H}^2_0}\ln \frac{k}{{\cal
H}_0}\right)^2  \sim  {k\over {\cal H}_0}.\label{P1}\ea Thus the
spectrum
is strongly blue-tilt. Here, it is just the superinflationary
evolution that brings the modulating mode with
$1-\epsilon_{Pre-inf}\gtrsim 1$ and $n_{\cal R}-1\gtrsim 1$ on
large angular scales. As showed in Eq.(\ref{tp}), the
corresponding mode will contribute a large modulation on the power
spectrum. Thus this model may result in the dipole power asymmetry
on corresponding scale, which is consistent with the observation
$A(k)\sim 0.07$. We plot ${\cal P}_{\cal R}$ in Eq.(\ref{ps}) in
Fig.1, which is consistent with our analytical result.

In Ref.\cite{Liu:2013kea}, a slightly similar spectrum has been
found for a bouncing inflation model, in which before the slow
roll inflation the Universe is in a contracting phase, see
Ref.\cite{Piao:2003zm} for an earlier study.

While at intermediate and small angular scales, i.e. $k \gg {\cal
H}_0$, we have \ba {\cal P}_{\cal R}^{k
> {\cal H}_0}\simeq
{\cal P}_{\cal R}^{\rm inf}\left(1-\frac{{3{\cal H}_0}}{{8k}}\sin
(\frac{2k}{{\cal H}_0})\right). \label{P2}\ea Thus the spectrum is
almost scale invariant but modulated with a small oscillation,
which is the standard result of slow-roll inflationary evolution.
Thus on corresponding scales the dipole power asymmetry is small,
which may be consistent with the constraint from the SDSS sample
of quasars \cite{Hirata} and also \cite{Flender:2013jja}.

\section{The CMB angular power spectrum with Planck}

\begin{figure}[t]
\begin{center}
\includegraphics[width=6cm]{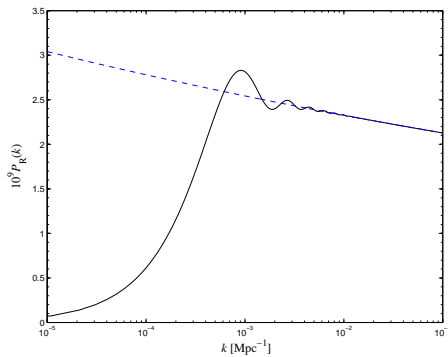}
\caption{Best-fit primordial power spectrum of curvature
perturbations for the pure power law (dashed) and our model
(solid) using Planck+WP data.}
\label{figpps}
\end{center}
\end{figure}

We will show the fit of our model to CMB TT spectrum, and also the
corresponding signals in TE and EE power spectra.

The slow-roll inflationary spectrum ${\cal P}_{\cal R}^{\rm inf}$
in Eq.(\ref{ps}) may be parameterized as a power law with ${\cal
P}_{\cal R}^{\rm inf}=A_{\rm inf}(k/k_0)^{n_{\rm inf}-1}$, where
$A_{\rm inf}$ is
the amplitude of perturbation, 
 see~\cite{Gauthier:2012aq} for possible features in the primordial
power spectrum and~\cite{guo05} for a general shape reconstructed
from CMB data. We follow Ref.\cite{Ade:2013nlj} and choose the pivot
scale to be $k_0=0.05 {\rm Mpc}^{-1}$, roughly in the middle of the
logarithmic range of scales probed by Planck.
We assume that the late-time cosmology is the standard flat
$\Lambda$CDM model described by four free cosmological parameters:
$\Omega_bh^2$, $\Omega_ch^2$, $\Theta_s$ and $\tau$. Here $h$ is
the dimensionless Hubble parameter such that $H_0$ = 100h
kms$^{-1}$ Mpc$^{-1}$ (noting that here $H_0$ is not related with
the cutoff scale ${\cal H}_0$, ), $\Omega_b h^2$ and $\Omega_ch^2$
are the physical baryon and dark matter densities relative to the
critical density at the present day, respectively, $\Theta_s$ is
the ratio of the sound horizon to the angular diameter distance at
the photon decoupling, and $\tau$ is the Thomson scattering
optical depth due to reionization.

\begin{figure}[t]
\begin{center}
\includegraphics[width=4cm]{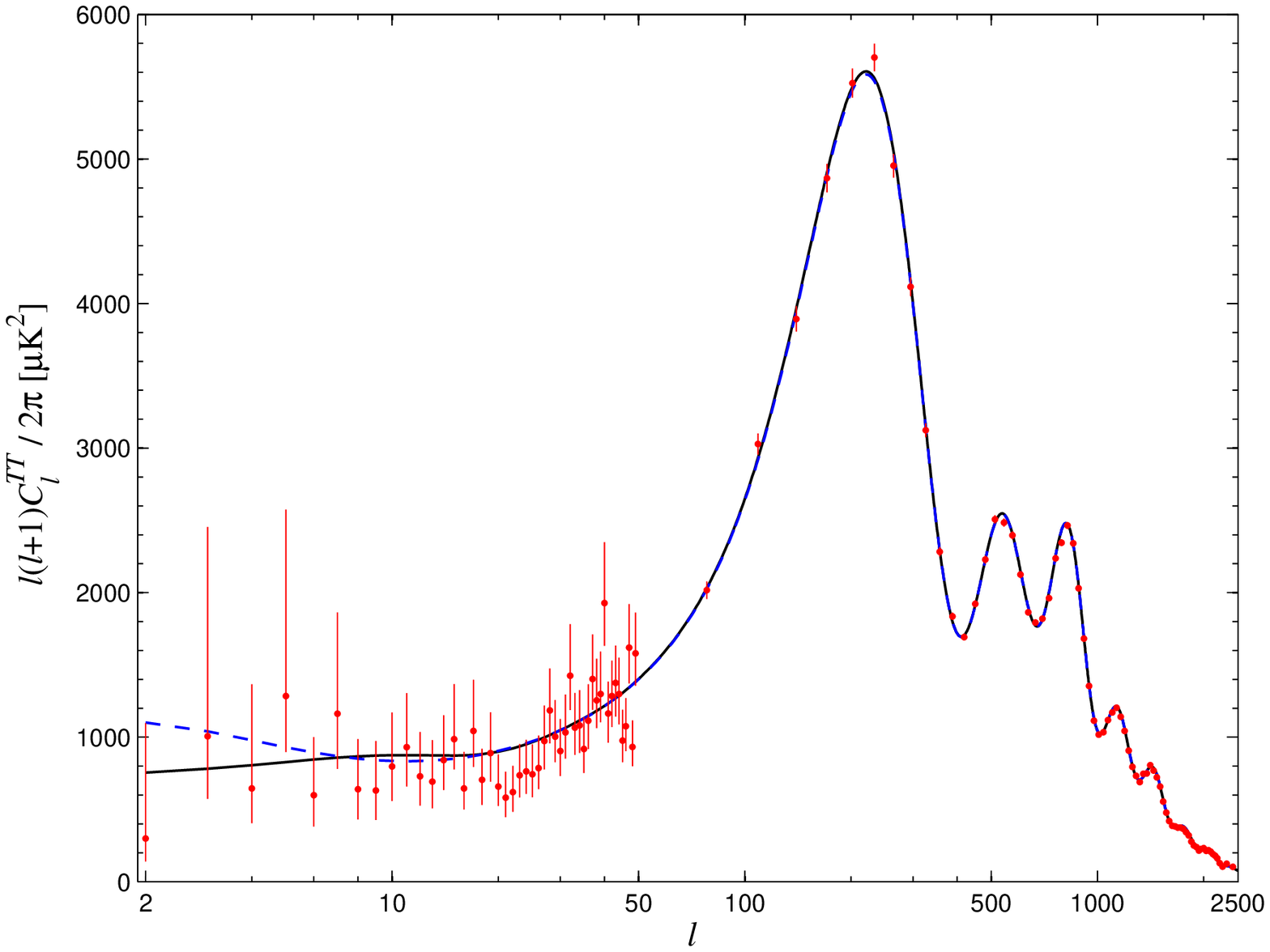}
\includegraphics[width=4cm]{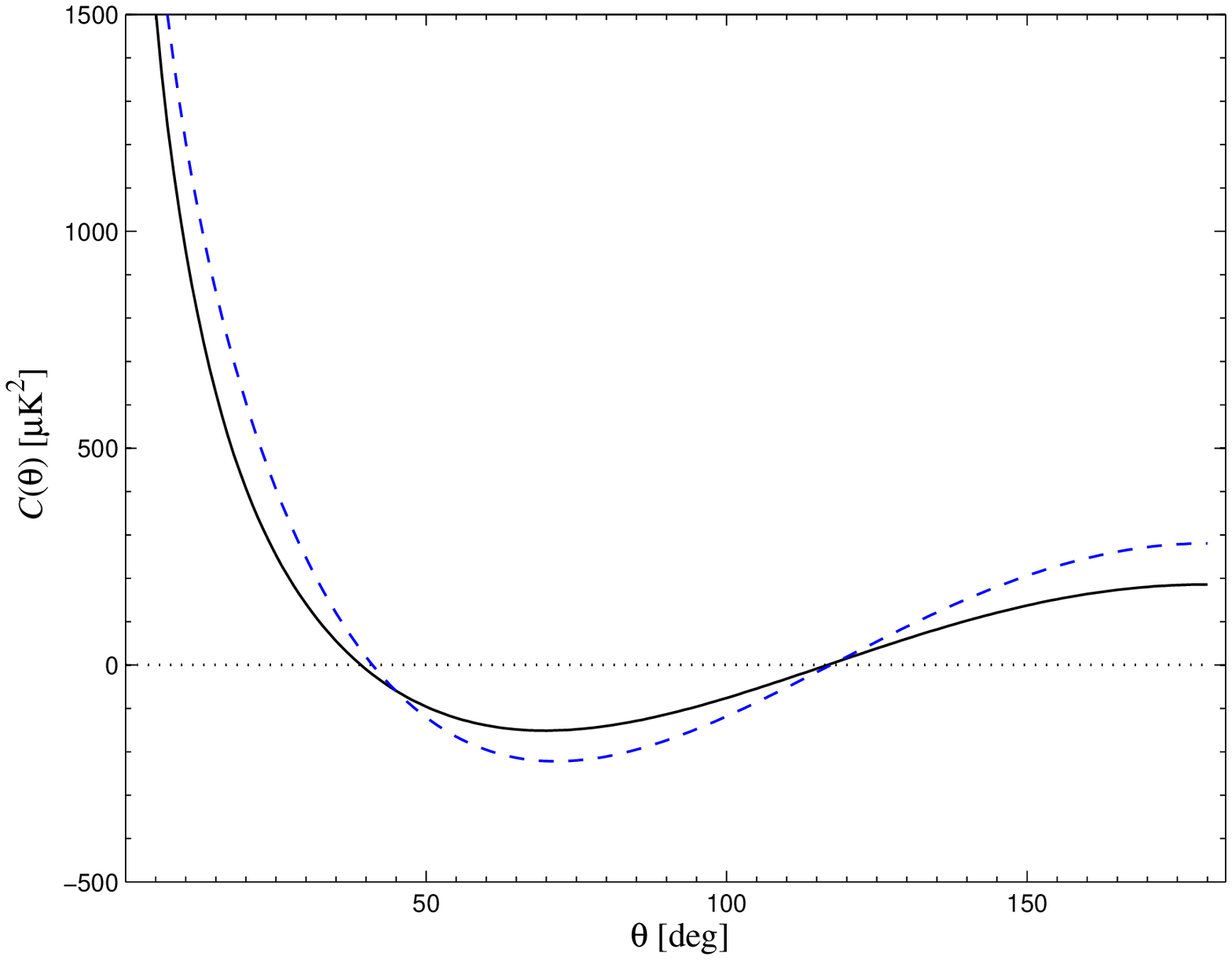}
\includegraphics[width=4cm]{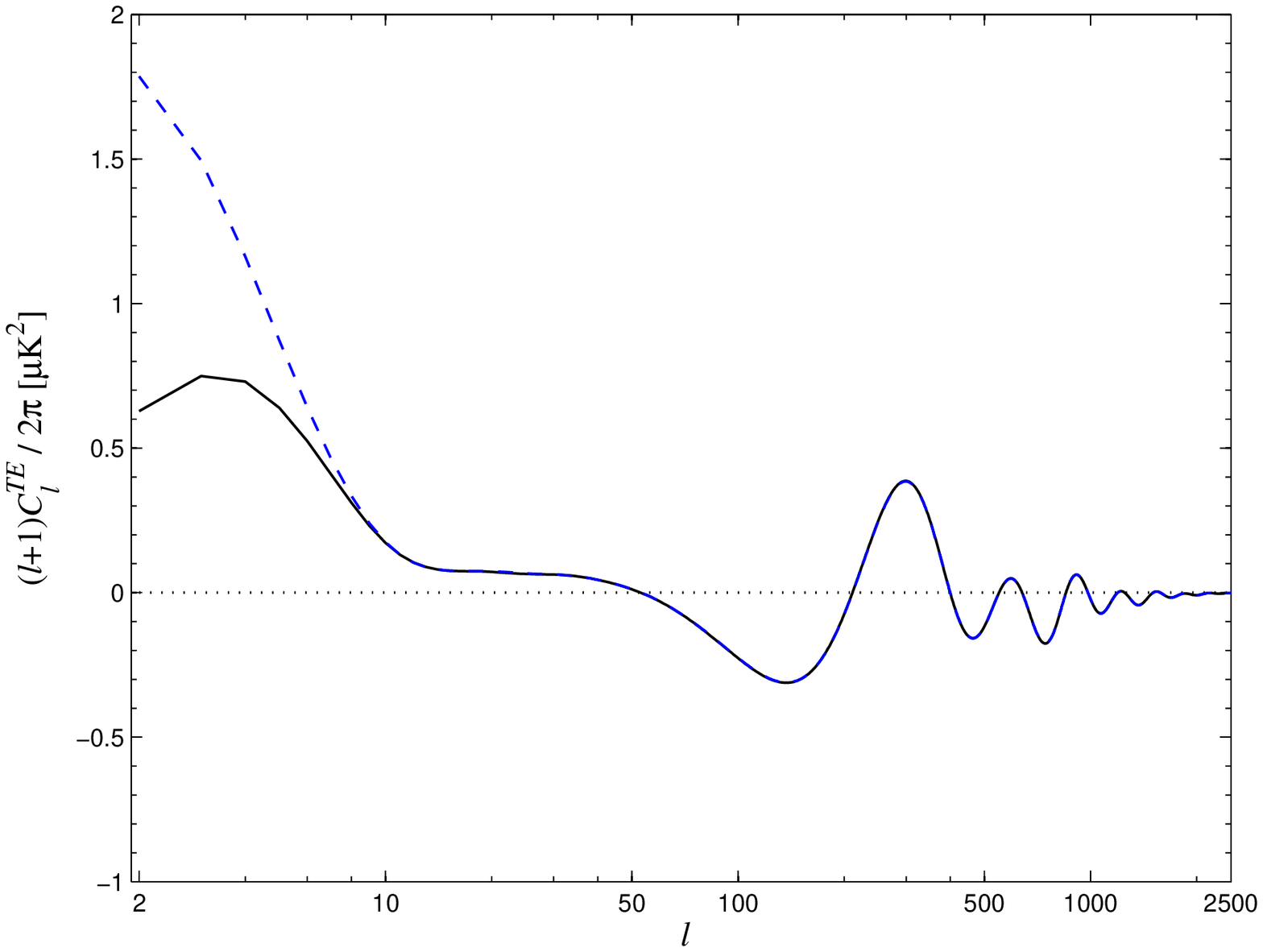}
\includegraphics[width=4cm]{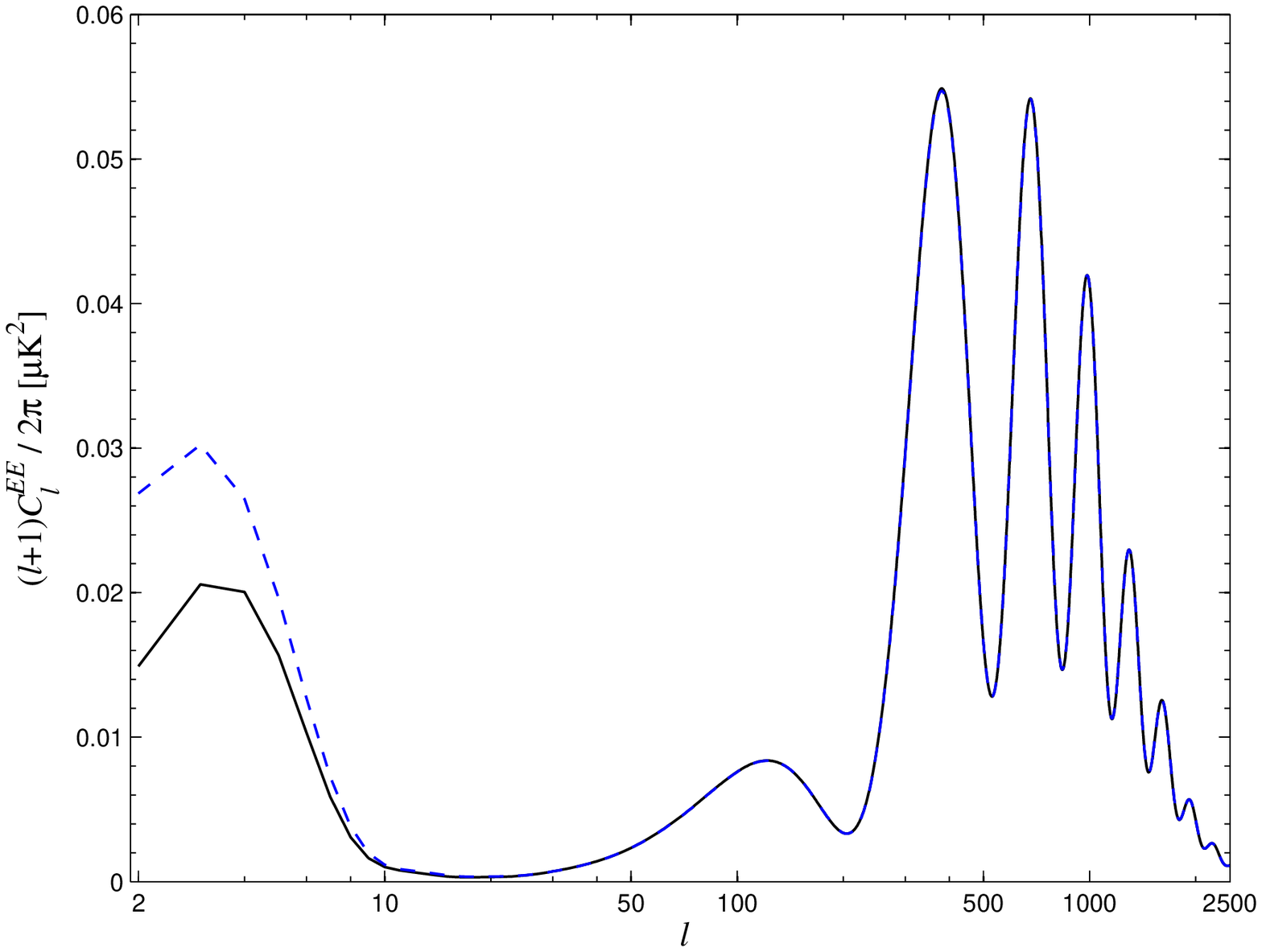}
\caption{Best-fit TT (upper left), TE (lower left), EE (lower right)
power spectra, and 2-point correlation function (upper right) for
the pure power law (dashed) and our model (solid) using Planck+WP data.
The red points show the Planck data with 1$\sigma$ errors.}
\label{figte}
\end{center}
\end{figure}

We modify the numerical Boltzmann code CAMB~\cite{Lewis:1999bs} to
calculate the lensed TT, TE, EE power spectra and 2-point
correlation function, and show the results in Fig.~\ref{figte}.
The blue dashed curves show the pure power law while the black
solid curves show our model~\eqref{ps} with the best-fit value of
$\ln ({\cal H}_0/{\rm Mpc}^{-1})=-7.47$. We see that the TT, TE
and EE spectra for our model are suppressed in the range $l<6$,
compared to the pure power law. Since the corresponding signals
are induced in the TE and EE spectra, the ongoing Planck
polarization data are expected to improve the constraints on the
model parameter ${\cal H}_0$. As shown in~\cite{Gruppuso:2010nd},
the polarization data can be used to test the parity asymmetry of
the CMB pattern. Note that there is a small bump around $l\sim 10$
in the TT spectrum due to oscillations of the primordial power
spectrum at large scales. The predicted 2-point correlation
function at $\theta > 50^\circ$ fits the Planck data much better
than the pure power-law spectrum~\cite{Copi:2013cya}.

We use the Planck CMB temperature likelihood~\cite{Ade:2013nlj}
supplemented by the WMAP large-scale polarization
likelihood~\cite{Page:2006hz} (Planck+WP). The Planck temperature
likelihood consists of the high-$l$ TT data ($50 \leq l \leq
2500$) and the low-$l$ TT data ($2 \leq l \leq 49$). Because of
contributions to the multi-frequency spectra from unresolved radio
point sources, cosmic infrared background, Sunyaev-Zeldovich
effects, calibration and beam uncertainties, the Planck high-$l$
likelihood includes 14 nuisance parameters which should be
marginalized in the analysis. As discussed in~\cite{Ade:2013nlj},
the large-scale E-mode polarization data is important for
constraining reionization. Hence we also use the 9-year WMAP
large-scale polarization likelihood including the TE, EE and BB
spectra in the range $2\leq l \leq 23$.

\begin{figure}[t]
\begin{center}
\includegraphics[width=6cm]{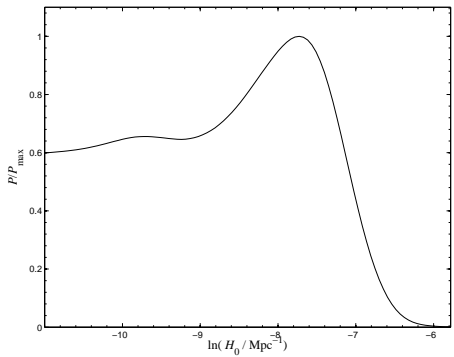}
\caption{Marginalized posterior distributions for ${\cal H}_0$
from the Planck+WP data.} \label{figkc}
\end{center}
\end{figure}

We use the Markov Chain Monte Carlo sampler as implemented in the
CosmoMC package~\cite{Lewis:2002ah} to construct the posterior
parameter probabilities. Since the Planck high-$l$ likelihood
includes many nuisance parameters which are fast parameters, a new
sampling method for decorrelating fast and slow parameters is
adopted in our analysis to efficiently scan the parameter
space~\cite{Lewis:2013hha}. We impose a flat prior on the
logarithm of ${\cal H}_0$ in the range [$-12,-4$]. For the other
cosmological parameters, prior ranges are chosen to be much larger
than the posterior. For the Planck+WP likelihood we find the
best-fit value of $\ln ({\cal H}_0/{\rm Mpc}^{-1})=-7.47$ with
$-2\ln{\cal L}_{\rm max}=9803.0$. This means that our model can
improve the fit to the data with $-2\Delta\ln{\cal L}_{\rm
max}=-4.8$ with respect to the standard power law model. However,
a two-parameter exponential-form cutoff of the primordial power
spectrum improves the fit only with $-2\Delta\ln{\cal L}_{\rm
max}=-2.9$ reported in~\cite{Ade:2013uln}. The reason is that the
small bump in the temperature spectrum induced by oscillation of
primordial power spectrum improves the fit to the data.
Fig.~\ref{figkc} shows the marginalized posterior distributions
for ${\cal H}_0$ from the Planck+WP data, which illustrates the
asymmetric shape of the likelihood functions.

Recently, some explanations appeared which attempted to provide a
mechanism to the anomalies,
\cite{Lyth:2013vha},\cite{Namjoo:2013fka},\cite{Kanno:2013ohv},\cite{Wang:2013lda},\cite{McDonald:2013aca},\cite{Cai:2013gma},\cite{Kohri:2013kqa},
and also \cite{D'Amico:2013iaa}. However, most of them involved
only the dipole power asymmetry in CMB, not the lack of power on
large angular scales. By contrast, our model not only generates
the power asymmetry but also a suppression of power on large
angular scales, see also \cite{Liu:2013kea} for a bouncing
inflationary model.

The power suppression on large angular scales has been also
implemented in fibre inflation
\cite{Cicoli:2013oba},\cite{Pedro:2013pba},\cite{Cicoli:2008gp},
and also \cite{Dudas:2012vv} for brane SUSY breaking models
\cite{Dudas:2010gi}, and \cite{Jain:2008dw} for the punctuated
inflation. However, how to explain the dipole power asymmetry in
the CMB was not illustrated in these studies.

\section{An implementation in string theory}

How embedding such an inflationary model into string theory is
interesting. We will discuss an implementation of our model in
string theory. In a warped compactifications with the brane/flux
annihilation \cite{KPV}, the effective potential controlling the
relevant evolution may potentially support a cosmological
inflation \cite{PRZ},\cite{DKV}. However, we find that there may
be a superinflationary phase before the slow-roll inflation.

In a 10 dimension CY manifold with the warped KS throat, the
metric of warped throat is \be ds^2= {1\over
\sqrt{f(r)}}ds^2_{(4)}+\sqrt{f(r)}(dr^2+r^2 ds_{(5)}^2) \ee for
$r<r_*$, where $r$ is the proper distance to the tip of the
throat, $ds^2_{(5)}$ is the angular part of the internal metric,
and $f(r)$ is the warp factor, which has a minimal value at $r_0$
and is determined by $\beta\equiv {r_0\over R}\sim e^{-{2\pi {\cal
K}\over 3g_s {\cal M}}}$, in which $R^4 ={27\pi\over 4} g_s
N\alpha^{\prime 2}$, $N$ equals to the product of the fluxes
${\cal M}$ and ${\cal K}$ for the RR and NSNS 3-forms,
respectively, $g_s$ is the string coupling and $\alpha^\prime$ is
set by the string scale. When $r>r_*$, this metric can be glued to
the metric of the bulk of the compact space, which is usually
taken to be a CY manifold. When $r_0< r <r_{*}$, $f(r)$ is
approximately $f(r)= ({R\over r})^4$.

\begin{figure}[t]
\begin{center}
\includegraphics[width=8cm]{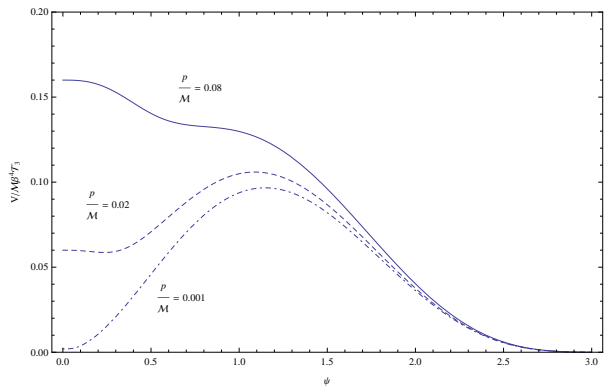}
\caption{The figure of the potential (\ref{vpsi}). When \bd-branes
are pulled into the throat continuously, the metastable minimum
will rise inch by inch.
} \label{figp}
\end{center}
\end{figure}

We follow Ref.\cite{KPV}. When $p(\ll {\cal M})$ \bd-branes sit at
the tip of KS throat, the system is a nonsupersymmetric NS5-brane
``giant graviton'' configuration, in which the NS5-brane warps a
$S^2$ in $S^3$, and carries $p$ unites flux, which induces the
\bd-charge. $S^2$ is inclined to expand as a spherical shell in
$S^3$, which may be parameterized by an angle $0 \leqslant \psi
\leqslant \pi $, in which $\psi =0$ corresponds to the north pole
of $S^3$ and $\psi =\pi$ is the south pole. The angular position
may be regarded as a scalar in the world volume action, which
describes the motion of the NS5-brane across the $S^3$. The
effective potential controlling the relevant evolution is \be
V_{\rm eff}(\psi)= {\cal M}\beta^4 T_3\left(\sqrt{{b_0^2
\sin^4{\psi}\over \pi^2}+{\tilde V}^2(\psi)}+{\tilde
V}(\psi)\right) \label{vpsi}\ee with $b_0\simeq 0.9$, where
${\tilde V}(\psi) ={p\over {\cal M}}-{\psi-{\sin{(2\psi)}\over
2}\over \pi}$ and $T_3$ is the \bd-brane tension. This potential
is plotted in Fig.~\ref{figp} with respect to $\psi$.

In the regime with $p/{\cal M} < 0.08$, the metastable bound state
forms, which corresponds to a static NS5-brane wrapping a $S^2$ in
$S^3$.
This metastable bound state corresponds to $\psi = 0$ and $ V_{\rm
eff}(0)=2p\beta^4 T_3$, see Fig.4. While the true minimum is at
$\psi = \pi$, in which the potential energy is $0$.

In the regime $p/{\cal M} \gtrsim 0.08$, this metastable state
disappears, which implies that the nonsupersymmetric configuration
of $p$ \bd-branes becomes classically unstable and will relax to a
supersymmetric minimum by a classical rolling of $\psi$ along its
potential. This classical rolling may lead to a slow-roll
inflation, which has been studied in detail in Ref.\cite{DKV}.
When $\psi =\pi$, in which the potential energy is $0$, the
inflation will end. The result of this evolution is ${\cal M}-p$
D3-branes instead of the original $p$ \bd-branes appearing at the
tip of throat, while the 3-form flux ${\cal K}$ is changed to
${\cal K}-1$, i.e. the brane/flux annihilation \cite{KPV}.

During the period before the slow roll inflation, in which
$p/{\cal M} < 0.08$, the Hubble expansion of the Universe is given
by \be H^2= {2p\beta^4 T_3 \over 3}, \label{h}\ee where
$8\pi/M_P^2= 1$. When \bd-branes are pulled into the throat
continuously, the metastable minimum will rise \cite{Piao:2007ne},
see Fig.~\ref{figp}, which implies that $H$ will increase rapidly
during this period.
Thus the parameter $\epsilon$ is \be \epsilon_{Pre-inf} =-{{\dot
H}\over H^2}\sim -\left({{\dot p}\over 2Hp}\right). \ee Thus in
unit of $\Delta t= 1/H$, we approximately have
$|\epsilon_{Pre-inf}| \sim {\Delta p\over 2p}$, where $\Delta p$
is the change of $p$ in unit of $1/H$.
We assume ${\Delta p\over 2p}\gtrsim 1$, which may be consistent
with ${\cal M}\sim 10^4$ and $p_{I}\sim {\cal O}(1)$, where
$p_{I}$ is the initial number of \bd-branes at the tip of the KS
throat. Here, all the moduli is assumed to be fixed, and the
interaction between \bd-branes has been also neglected for
simplicity.


Thus in this model the Universe initially is in a
superinflationary phase with $\epsilon_{Pre-inf}\sim -{\cal
O}(1)$, during which the number of \bd-branes at the tip of throat
will increase rapidly. After a sufficient number of \bd-branes
enter into the throat, which makes $p$ reaching its critical
value, $\psi$ will slowly roll down to its real minimum at $\psi=
\pi$, during which the Universe is in a slow-roll inflationary
phase. Thus as has been argued, it is just the stringy physics
before the slow-roll inflation that results in a large-scale
cutoff in the primordial power spectrum.

We conclude that a stringy model of inflation
in which initially the Universe is in a superinflationary phase,
can generate a large-scale cutoff in the primordial power
spectrum, which
may account for not only the power suppression on large angular
scales, but also a large dipole power asymmetry in the CMB. In the
meantime this model also predicts distinct signals in TE and EE
power spectra, which may be falsified by the observation of CMB
polarization.


\textbf{Acknowledgments}

We thank Tirthabir Biswas, David H. Lyth, Anupam Mazumdar for
helpful discussions. ZKG is supported by the project of Knowledge
Innovation Program of Chinese Academy of Science, NSFC under Grant
No.11175225, and National Basic Research Program of China under
Grant No.2010CB832805. YSP is supported by NSFC under Grant
No.11075205, 11222546, and National Basic Research Program of
China, No.2010CB832804. We used CosmoMC and CAMB. We acknowledge
the use of the Planck data and the Lenovo DeepComp 7000
super-computer in SCCAS.


\begin{thebibliography}{99}



\bibitem{Ade:2013nlj}
  P.~A.~R.~Ade {\it et al.}  [Planck Collaboration],
  arXiv:1303.5083.



\bibitem{Paci:2010wp}
  F.~Paci, A.~Gruppuso, F.~Finelli, P.~Cabella, A.~De Rosa, N.~Mandolesi and P.~Natoli,
  Mon.\ Not.\ Roy.\ Astron.\ Soc.\  {\bf 407}, 399 (2010)
  [arXiv:1002.4745].

\bibitem{Dai:2013kfa}
  L.~Dai, D.~Jeong, M.~Kamionkowski and J.~Chluba,
  Phys.\ Rev.\ D {\bf 87}, 123005 (2013)  [arXiv:1303.6949
  [astro-ph.CO]].

\bibitem{Liddle:2013czu}
  A.~R. Liddle and M.~Cort¨ºs,
  Phys.\ Rev.\ Lett.\  {\bf 111}, 111302 (2013)
  [arXiv:1306.5698].

\bibitem{Prunet:2004zy}
  S.~Prunet, J.~P.~Uzan, F.~Bernardeau and T.~Brunier,
  Phys.\ Rev.\ D {\bf 71}, 083508 (2005)
  [astro-ph/0406364].

\bibitem{Gordon:2005ai}
  C.~Gordon, W.~Hu, D.~Huterer and T.~M.~Crawford,
  Phys.\ Rev.\ D {\bf 72}, 103002 (2005)
  [astro-ph/0509301].

\bibitem{Rath:2013yra}
  P.~K.~Rath and P.~Jain,
  arXiv:1308.0924.



\bibitem{Erickcek:2008sm}
  A.~L.~Erickcek, M.~Kamionkowski and S.~M.~Carroll,
  Phys.\ Rev.\ D {\bf 78}, 123520 (2008)
  [arXiv:0806.0377].

\bibitem{Lyth:2013vha}
  D.~H.~Lyth,
  JCAP {\bf 1308}, 007 (2013)  [arXiv:1304.1270 [astro-ph.CO]].



\bibitem{Liu:2013kea}
  Z.~G.~Liu, Z.~K.~Guo and Y.~S.~Piao,
  Phys.\ Rev.\ D {\bf 88}, 063539 (2013)
  [arXiv:1304.6527].







\bibitem{Namjoo:2013fka}
  M.~H.~Namjoo, S.~Baghram and H.~Firouzjahi,
  Phys.\ Rev.\ D {\bf 88}, 083527 (2013)  [arXiv:1305.0813 [astro-ph.CO]];
  A.~A.~Abolhasani, S.~Baghram, H.~Firouzjahi and M.~H.~Namjoo,
  arXiv:1306.6932.

\bibitem{Piao:2003zm}
  Y.~S.~Piao, B.~Feng and X.~m.~Zhang,
  Phys.\ Rev.\ D {\bf 69}, 103520 (2004)
  [hep-th/0310206];
  Y.~S.~Piao,
  Phys.\ Rev.\ D {\bf 71}, 087301 (2005)
  [astro-ph/0502343].


\bibitem{Dudas:2012vv}
  E.~Dudas, N.~Kitazawa, S.~P.~Patil and A.~Sagnotti,
  JCAP {\bf 1205} (2012) 012; 
  C.~Condeescu and E.~Dudas,
  JCAP {\bf 1308} (2013) 013.


\bibitem{Burgess:2013sla}
  C.~P.~Burgess, M.~Cicoli and F.~Quevedo,
  JCAP {\bf 1311}, 003 (2013).


\bibitem{Kanno:2013ohv}
  S.~Kanno, M.~Sasaki and T.~Tanaka,
  arXiv:1309.1350.

\bibitem{Cicoli:2013oba}
  M.~Cicoli, S.~Downes and B.~Dutta,
  arXiv:1309.3412 [hep-th].

\bibitem{Pedro:2013pba}
  F.~G.~Pedro and A.~Westphal,
  arXiv:1309.3413 [hep-th].

\bibitem{Cicoli:2008gp}
  M.~Cicoli, C.~P.~Burgess and F.~Quevedo,
  JCAP {\bf 0903}, 013 (2009)  [arXiv:0808.0691 [hep-th]].


\bibitem{PRZ}
  L.~Pilo, A.~Riotto and A.~Zaffaroni,
  JHEP {\bf 0407}, 052 (2004)
  [hep-th/0401004].


\bibitem{DKV}
  O.~DeWolfe, S.~Kachru and H.~L.~Verlinde,
  JHEP {\bf 0405}, 017 (2004)
  [hep-th/0403123].






\bibitem{Piao:2004tq}
  Y.~S.~Piao and Y.~Z.~Zhang,
  Phys.\ Rev.\ D {\bf 70}, 063513 (2004)
  [astro-ph/0401231];

\bibitem{Piao:2003ty}
  Y.~S.~Piao and E.~Zhou,
  Phys.\ Rev.\ D {\bf 68}, 083515 (2003)
  [hep-th/0308080].

\bibitem{Piao:2010bi}
  Y.~-S.~Piao,
  Phys.\ Lett.\ B {\bf 701}, 526 (2011)  [arXiv:1012.2734
  ]; 
  Z.~-G.~Liu, J.~Zhang and Y.~-S.~Piao,
  Phys.\ Rev.\ D {\bf 84}, 063508 (2011)  [arXiv:1105.5713
  ]; 
  Z.~-G.~Liu and Y.~-S.~Piao,
  Phys.\ Lett.\ B {\bf 718}, 734 (2013)  [arXiv:1207.2568 ].


\bibitem{Biswas:2013dry}
  T.~Biswas and A.~Mazumdar,
  arXiv:1304.3648 [hep-th].










\bibitem{Hirata}
  C. M. Hirata,
  JCAP {\bf 0909}, 011 (2009)
  [arXiv:0907.0703].

\bibitem{Flender:2013jja}
  S.~Flender and S.~Hotchkiss,
  JCAP {\bf 1309}, 033 (2013)
  [arXiv:1307.6069].


\bibitem{Gauthier:2012aq}
  C.~Gauthier and M.~Bucher,
  JCAP {\bf 1210}, 050 (2012)
  [arXiv:1209.2147].
\bibitem{guo05}
  Z.~K.~Guo, D.~J.~Schwarz and Y.~Z.~Zhang,
  JCAP {\bf 1108}, 031 (2011)
  [arXiv:1105.5916];
  Z.~K.~Guo and Y.~Z.~Zhang,
  JCAP {\bf 1111}, 032 (2011)
  [arXiv:1109.0067];
  Z.~K.~Guo and Y.~Z.~Zhang,
  Phys.\ Rev.\ D {\bf 85}, 103519 (2012)
  [arXiv:1201.1538].
\bibitem{Lewis:1999bs}
  A.~Lewis, A.~Challinor and A.~Lasenby,
  Astrophys.\ J.\  {\bf 538}, 473 (2000)
  [astro-ph/9911177].
\bibitem{Gruppuso:2010nd}
  A.~Gruppuso, F.~Finelli, P.~Natoli, F.~Paci, P.~Cabella, A.~De Rosa and N.~Mandolesi,
  Mon.\ Not.\ Roy.\ Astron.\ Soc.\  {\bf 411}, 1445 (2011)
  [arXiv:1006.1979].
\bibitem{Copi:2013cya}
  C.~J.~Copi, D.~Huterer, D.~J.~Schwarz and G.~D.~Starkman,
  Adv.\ Astron.\  {\bf 2010}, 847541 (2010)
  [arXiv:1004.5602];
  C.~J.~Copi, D.~Huterer, D.~J.~Schwarz and G.~D.~Starkman,
  arXiv:1310.3831.
\bibitem{Page:2006hz}
  L.~Page {\it et al.}  [WMAP Collaboration],
  Astrophys.\ J.\ Suppl.\  {\bf 170}, 335 (2007)
  [astro-ph/0603450].
\bibitem{Lewis:2002ah}
  A.~Lewis and S.~Bridle,
  Phys.\ Rev.\ D {\bf 66}, 103511 (2002)
  [astro-ph/0205436].
\bibitem{Lewis:2013hha}
  A.~Lewis,
  ``Efficient sampling of fast and slow cosmological parameters,''
  Phys.\ Rev.\ D {\bf 87}, 103529 (2013)
  [arXiv:1304.4473].
\bibitem{Ade:2013uln}
  P.~A.~R.~Ade {\it et al.}  [Planck Collaboration],
  arXiv:1303.5082.


\bibitem{Wang:2013lda}
  L.~Wang and A.~Mazumdar,
  Phys.\ Rev.\ D {\bf 88}, 023512 (2013)  [arXiv:1304.6399
  [astro-ph.CO]].



\bibitem{McDonald:2013aca}
  J.~McDonald,
  JCAP {\bf 1307}, 043 (2013)  [arXiv:1305.0525 [astro-ph.CO]];
  J.~McDonald,
  arXiv:1309.1122 [astro-ph.CO].




\bibitem{Cai:2013gma}
  Y.~-F.~Cai, W.~Zhao and Y.~Zhang,
  arXiv:1307.4090 [astro-ph.CO].




\bibitem{Kohri:2013kqa}
  K.~Kohri, C.~M.~Lin and T.~Matsuda,
  arXiv:1308.5790.


\bibitem{D'Amico:2013iaa}
  G.~D'Amico, R.~Gobbetti, M.~Kleban and M.~Schillo,
  arXiv:1306.6872 [astro-ph.CO].



\bibitem{Dudas:2010gi}
  E.~Dudas, N.~Kitazawa and A.~Sagnotti,
  Phys.\ Lett.\ B {\bf 694} (2010) 80
  [arXiv:1009.0874 [hep-th]]; 
  A.~Sagnotti,
  arXiv:1303.6685 [hep-th]; 
  P.~Fr$\acute{e}$, A.~Sagnotti and A.~S.~Sorin,
  arXiv:1307.1910 [hep-th].


\bibitem{Jain:2008dw}
  R.~K.~Jain, P.~Chingangbam, J.~-O.~Gong, L.~Sriramkumar and T.~Souradeep,
  JCAP {\bf 0901}, 009 (2009)  [arXiv:0809.3915 [astro-ph]]; 
  R.~K.~Jain, P.~Chingangbam, L.~Sriramkumar and T.~Souradeep,
  Phys.\ Rev.\ D {\bf 82}, 023509 (2010)  [arXiv:0904.2518
  [astro-ph.CO]].

\bibitem{KPV}
  S.~Kachru, J.~Pearson and H.~L.~Verlinde,
  JHEP {\bf 0206}, 021 (2002)
  [hep-th/0112197].




\bibitem{Piao:2007ne}
  Y.~S.~Piao,
  Phys.\ Rev.\ D {\bf 78}, 023518 (2008)
  [arXiv:0712.3328].




\end{thebibliography}
\end{document}